\documentclass[aps,prl,preprint,superscriptaddress]{revtex4-2}

\usepackage{graphicx}%
\usepackage{amsmath,amssymb,amsfonts}%

\newcommand\lihox{LiHo\textsubscript{x}Y\textsubscript{1-x}F\textsubscript{4}}
\newcommand\lihoforty{LiHo\textsubscript{0.4}Y\textsubscript{0.6}F\textsubscript{4}}
\newcommand\offdiag{$V_{ij}^{zx} \sigma_i^z \sigma_j^x$}
\newcommand\diag{$V_{ij}^{zz} \sigma_i^z \sigma_j^z$}

\begin{document}

\title{Quantum Barkhausen Noise Induced by Domain Wall Co-Tunneling}
\author{C. Simon}
\author{D.M. Silevitch}
\affiliation{Division of Physics, Mathematics, and Astronomy, California Institute of Technology, Pasadena, CA 91125, USA}
\author{P.C.E. Stamp}
\affiliation{Division of Physics, Mathematics, and Astronomy, California Institute of Technology, Pasadena, CA 91125, USA}
\affiliation{Department of Physics and Astronomy, University of British Columbia, Vancouver, BC V6T 1Z1, Canada}
\affiliation{Pacific Institute of Theoretical Physics, University of British Columbia, Vancouver, BC V6T 1Z1, Canada}
\author{T.F. Rosenbaum}
\affiliation{Division of Physics, Mathematics, and Astronomy, California Institute of Technology, Pasadena, CA 91125, USA}

\begin{abstract}Most macroscopic magnetic phenomena (including magnetic hysteresis) are typically understood classically. Here, we examine the dynamics of a uniaxial rare-earth ferromagnet deep within the quantum regime, so that domain wall motion, and the associated hysteresis, is dominated by large-scale quantum tunneling of spins, rather than classical thermal activation over a potential barrier. The domain wall motion is found to exhibit avalanche dynamics, observable as an unusual form of Barkhausen noise \cite{barkhausen_notitle_1919}. We observe non-critical behavior in the avalanche dynamics that only can be explained by going beyond traditional renormalization group methods\cite{dahmen_hysteresis_1996} or classical domain wall models\cite{alessandro_domainwall_1990}. We find that this “quantum Barkhausen noise” exhibits two distinct mechanisms for domain wall movement, each of which is quantum-mechanical, but with very different dependences on an external magnetic field applied transverse to the spin (Ising) axis. These observations can be understood in terms of the correlated motion of pairs of domain walls, nucleated by co-tunneling of plaquettes (sections of domain wall), with plaquette pairs correlated by dipolar interactions; this correlation is suppressed by the transverse field. Similar macroscopic correlations may be expected to appear in the hysteresis of other systems with long-range interactions.\end{abstract}

\maketitle

\section{Introduction}

Although magnetism at the microscopic scale has been understood as a quantum phenomenon for nearly a century, macroscopic magnetic objects like domain walls are usually treated classically \cite{stamp_quantum_1992,hong_approach_1995,hong_effect_1996,wernsdorfer_quantum_2005,brooke_tunable_2001,silevitch_switchable_2010,silevitch_magnetic_2019}. There is good reason for this: in a conductor, the dissipative coupling to electrons rapidly suppresses domain wall tunneling \cite{tatara_macroscopic_1994,tatara_macroscopic_1994-1}, and even in an insulator, the coupling to phonons \cite{stamp_quantum_1991,dube_effects_1998}, paramagnetic impurities \cite{dube_effects_1998}, and nuclear spins \cite{dube_effects_1998} is enough to render the wall motion classical, except at microscopic scales. These mechanisms also suppress ``chiral tunneling'' \cite{takagi_macroscopic_1996,galkina_chirality_2008,dube_effects_1998} between opposite chiralities for a given wall; for a Bloch wall, the chirality is simply the sense (clockwise or anti-clockwise) in which the magnetization winds in passing between the states on either side of the wall.

It is actually very difficult to observe the dynamics of individual domain walls except in restricted geometries. More typically, one sees evidence of collective ``multi-wall'' motion, either by imaging walls before and after this motion has occurred, or by measurements of the dynamic susceptibility $\chi(\omega)$, or of the ``Barkhausen noise'' in the bulk magnet. The latter shows up in inductive measurements, arising from rapid jumps in the magnetization caused by the depinning of walls and their subsequent motion. Since the discovery of Barkhausen noise in 1919 \cite{barkhausen_notitle_1919}, a vast corpus of experimental work has accumulated, characterizing the influence on the noise of disorder, different magnetic interactions, and the proximity to phase transitions \cite{durin_barkhausen_2006,bohn_playing_2018,de_sousa_waiting-time_2020}. However, all of this work has been done on thermally-activated wall motion--there have been no investigations of quantum Barkhausen noise, in which domain wall tunneling, rather than thermal excitation over barriers, dominates.  

The present study explores this quantum regime. Just as in the classical regime, we expect inter-wall interactions, mediated by dipolar interactions, to lead to collective wall dynamics and even avalanche processes under the right conditions. We also expect the coupling to phonons to play a role.

To disentangle the different processes, it is important to choose the right experimental system. We need the wall structure to be simple, and to see quantum behavior, we need the crossover between classical thermal activation and quantum tunneling to be at sufficiently high temperature. An ideal system is the Ising magnet \lihox, in which very strong crystal fields acting on the J=8 Ho electronic spins create a low-temperature Ising doublet with renormalized moment $\Tilde{J} \approx 5.51 \mu_B$, separated from the 1st excited state by a gap $\Delta \approx 9.4 $K.

When T $\ll \Delta$, the system can be described by an effective Hamiltonian $\mathcal{H} = \mathcal{H}_{\text{QI}} + \mathcal{H}_{\text{env}}$, where $\mathcal{H}_{\text{QI}}$ is the quantum Ising Hamiltonian:
\begin{equation}
    \mathcal{H}_{\text{QI}} = - \sum_{i \neq j} V_{ij}^{zz} \sigma_i^z \sigma_j^z - \Gamma_0(B_x) \sum_i \sigma_i^x
\end{equation}
with a longitudinal dipolar interaction $V_{ij}^{zz}$ and a bare splitting $\Gamma_0(B_x)$ induced in the Ising doublet by a transverse field $B_x$. The ``environmental'' term $\mathcal{H}_{\text{env}}$ has the form
\begin{equation}
    \mathcal{H}_{\text{env}} = \mathcal{H}_{\text{hyp}} + \mathcal{H}_{\text{ph}} + \mathcal{H}_{\text{EM}} + \mathcal{H}_{\text{disorder}}
\end{equation}
where the first three terms refer to hyperfine interactions, spin-phonon interactions, and interactions with the electromagnetic field, and the final term describes the effect of disorder when the concentration of Holmium $x < 1$. For details of this Hamiltonian and its derivation, see supplementary information.

The main effect of $\mathcal{H}_{\text{hyp}}$ is to block flips between the $| \uparrow \rangle$ and $| \downarrow \rangle$ states of the Ising doublet \cite{schechter_significance_2005,schechter_derivation_2008} until $B_x \sim 2 \text{ T}$, giving a much reduced effective splitting $\Tilde{\Gamma}(B_x)$. The spin-phonon terms facilitate irreversible phonon-assisted flips, and $\mathcal{H}_{\text{EM}}$ contains the demagnetization field generated by the spins, which depends on sample shape. Finally, when $x < 1$, off-diagonal dipolar terms $\sim$ \offdiag are generated \cite{schechter_significance_2005,schechter_derivation_2008,schechter_quantum_2006}, so that an applied longitudinal field generates a random transverse field, and an applied transverse field generates a random longitudinal field.

Very thin domain walls (with thickness $\lambda < a_0$, the lattice spacing) are allowed because the exchange interaction between spins is negligible. The easy axis $\hat{z}$ lies in the wall plane and the wall orientation is controlled by pinning forces and the demagnetization fields. As noted many years ago by Egami \cite{egami_theory_1973-1,egami_theory_1973}, domain wall motion for such walls involves the nucleation and growth of ``plaquette'' distortions of the wall, in which a small section of the wall shifts locally by a single lattice spacing. In the presence of an external field along the easy $z$-axis, the plaquette can tunnel through a barrier created by the energy associated with the wall distortion, and then grow once it exceeds a critical size. This nucleation also requires a local transverse field to flip individual spins (otherwise $[\mathcal{H}, \sigma_i^z] = 0$). Such a field can be applied come from transverse demagnetization, or arise from the disorder-induced terms $\sim \sigma_i^z \sigma_j^x$ noted above.

Domain walls previously have been imaged in the \lihox\ system \cite{jorba_shpm_2014,meyer_magnetooptic_1989}, and indirect evidence for tunneling motion has been found in low-temperature susceptibility measurements in transverse magnetic field \cite{brooke_tunable_2001,silevitch_switchable_2010,silevitch_magnetic_2019}. A priori we expect the crossover between thermal activation and tunneling to occur at a relatively high temperature compared to typical Bloch wall systems because the wall is so thin, because very small regions are involved in plaquette formation, and because the characteristic frequency of spin flips is high. Taken together, these characteristics make the \lihox\ system a good candidate for low-temperature tunneling behavior.

\section{Results}

\begin{figure}[tb]
    \centering
    \includegraphics[width=0.4\textwidth]{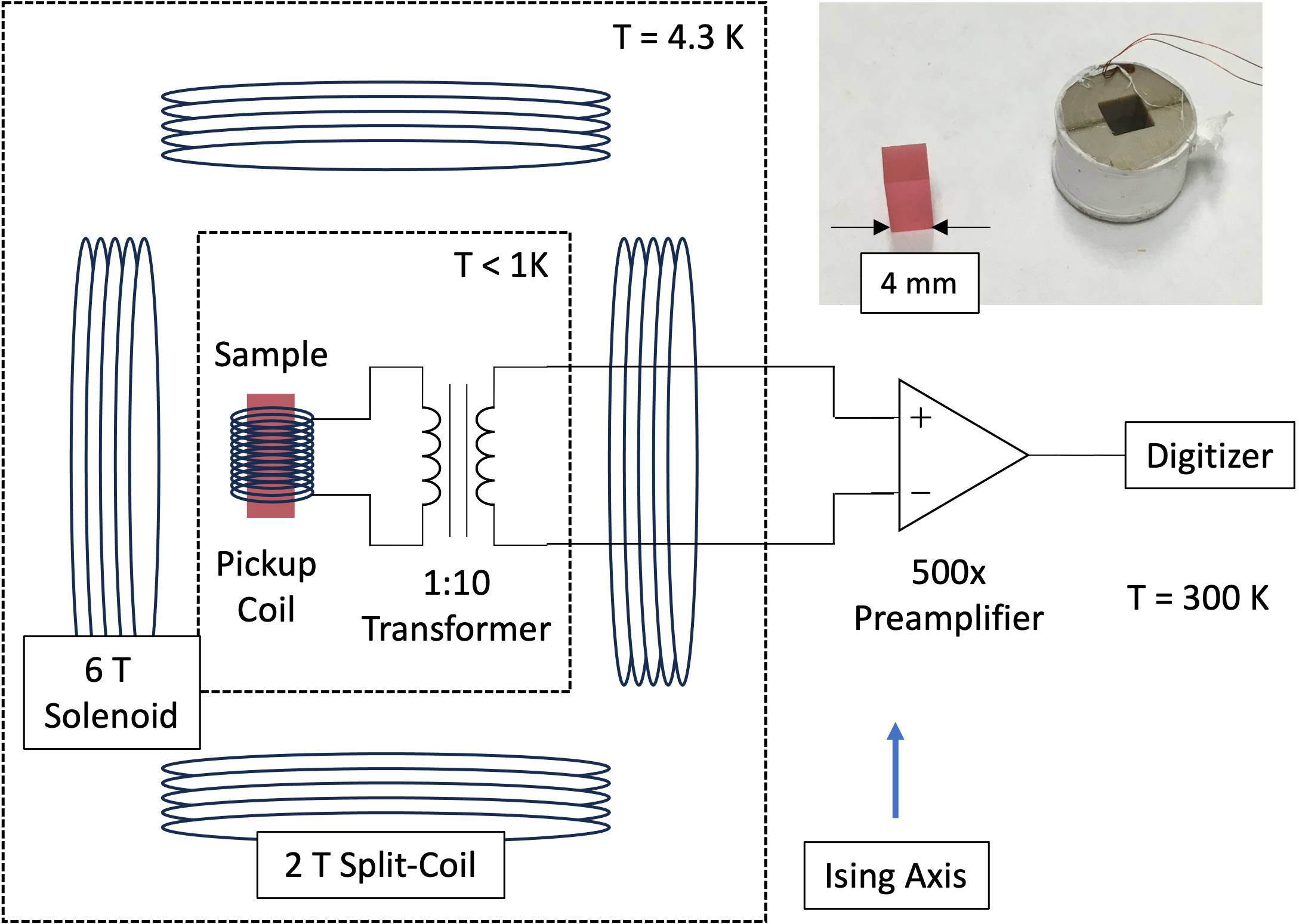}
    \caption{\textbf{Schematic of experimental setup} An inductive pickup coil is wound around the crystal of LiHo$_{0.40}$Y$_{0.60}$F${}_4$  inside an insulating PEEK coil form. The assembly is mounted on the Cu cold finger of a helium dilution refrigerator equipped with a 6T/2T superconducting vector magnet. The induced voltage signal is amplified first by a cryogenic broadband transformer amplifier, and then at room temperature by a low-noise transistor preamplifler  and finally digitized with a streaming oscilloscope. Inset: Photograph of sample and pickup coil assembly.}
    \label{fig:1}
\end{figure}

In the experiments described here, we made both quasistatic bulk magnetization measurements and faster time-domain measurements of individual magnetic avalanches known as ``Barkhausen events'' \cite{barkhausen_notitle_1919} on a single crystal of  \lihoforty\ system at temperatures ranging from 15 \% (90 mK) to 95 \% (580 mK) of the Curie temperature, $T_C = 612 \text{ mK}$, as the external longitudinal field (along the Ising axis) was ramped between  $H^\parallel = \pm 4 \text{ kOe}$--a field sufficiently large to saturate the magnetization. Additionally, a static external field transverse to the Ising axis was applied at several values ranging from $H^{\perp} = 0 \rightarrow 200 \text{ Oe}$. We emphasize that, to our knowledge, this is the first measurement of quantum Barkhausen noise. A diagram of the experimental setup is shown in Fig. \ref{fig:1}.

\begin{figure}[tb]
    \centering
    \includegraphics[width=0.4\textwidth]{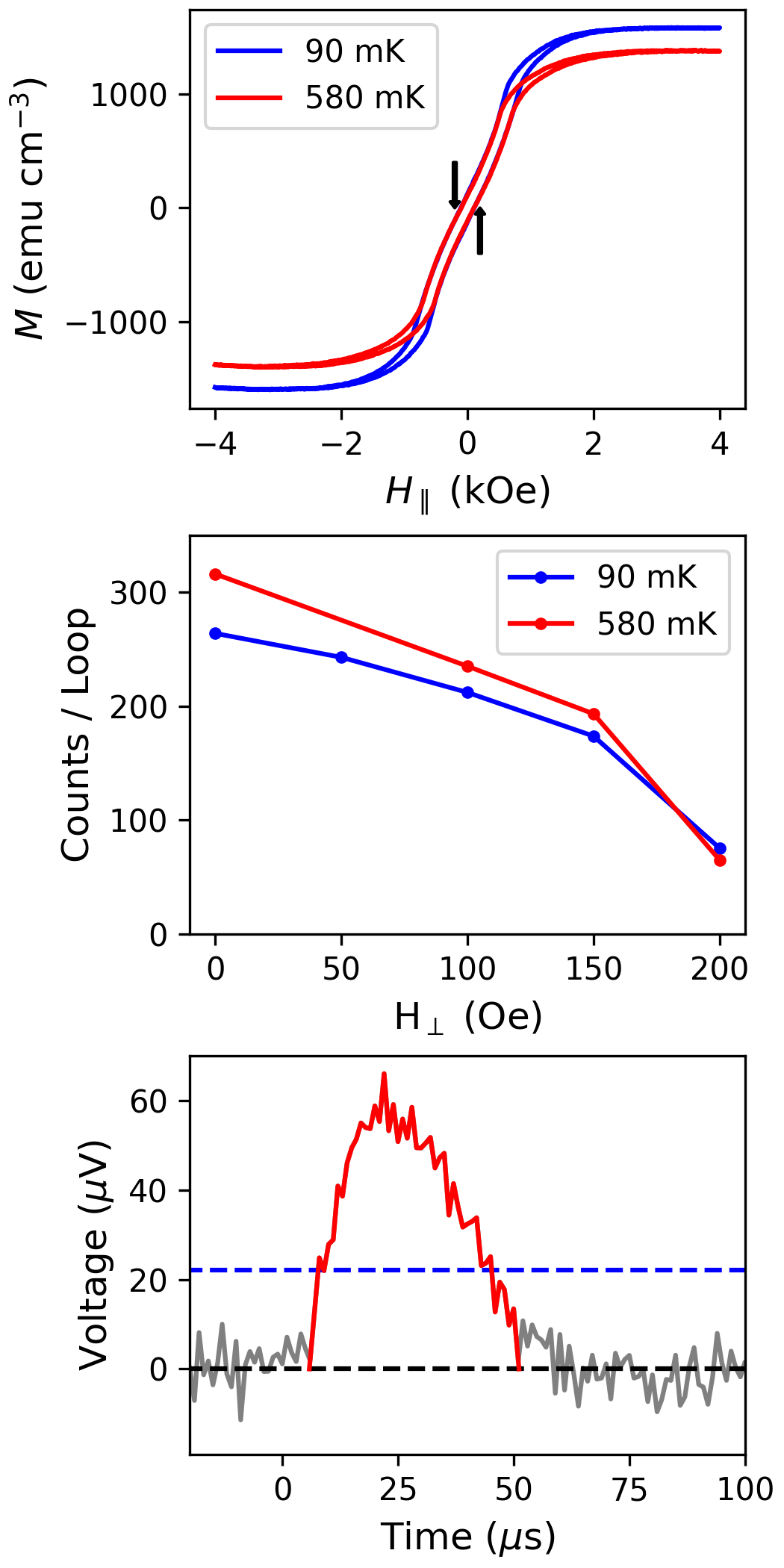}
    \caption{\textbf{Hysteresis loops and Barkhausen events for extremal temperatures.} \textbf{a}  Magnetization vs longitudinal field curves at temperatures 15\% and 95\% of the Curie temperature with $H_\perp = 0$.Loops at transverse fields up to 200 Oe are indistinguishable from those at zero transverse field. \textbf{b} Average number of detected events per loop for a given temperature and transverse field. \textbf{c} Event extraction for a sample event. Data in grey are the raw extracted voltages, blue is the detection threshold, and red is the extracted event}
    \label{fig:2}
\end{figure}

When comparing the DC bulk magnetization of the sample at different temperatures (Fig. \ref{fig:2}a), it is apparent that, at a macroscopic level, the effect of increasing temperature is to decrease the saturation magnetization, affecting the outer portions of the hysteresis loop at fields $|H_\parallel| > 1 \text{ kOe}$, but leaving the ``linear'' regime of the hysteresis loop unchanged at lower fields $|H_\parallel| < 1 \text{ kOe}$. Furthermore, the strength of the transverse fields applied in these measurements ($H_\perp \leq 200 \text{ Oe}$) was sufficiently weak that they had no observable effect on any portion of the hysteresis loop.

In many soft ferromagnets, the creation/annihilation of individual domains is found to occur at external longitudinal fields close to saturation\cite{herpin_theorie_nodate}, while the ``linear'' regime at lower fields is dominated by the motion of domain walls. Since we are primarily interested in the dynamics of the motion of domain walls, we restrict our analysis to this ``linear'' regime at low fields ( $|H_\parallel| < 600 \text{ Oe}$).

Even from 15 \% to 95 \% of $T_C$, there is no change in the sample response (Fig. \ref{fig:2}a), indicating that thermal fluctuations are unimportant in the dynamics of the domain wall motion. This implies that for all temperatures and fields the domain wall dynamics are driven by quantum, rather than thermal, fluctuations. This is to be contrasted with a previous measurement \cite{silevitch_switchable_2010} on a sample of similar Ho concentration $x=0.44$ (as opposed to our $x=0.4$), in which, under no transverse field, the hysteresis loop narrowed considerably from low (100 mK) to high (500 mK) temperature.  The essential difference between these two measurements is the sample shape, with the crystal here a cuboid with aspect ratio 1x1x2, vs. the previous measurement performed on a needle with a much longer aspect ratio. The difference in aspect ratios results in different shape-dependent profiles of the demagnetization field, explaining the discrepancy in the behavior. In particular, when combined with quenched site disorder (from chemical dilution of Ho by Y), the Ising spins not only exert spatially-varying longitudinal fields on each other through the diagonal dipolar interaction \diag, but also spatially-varying transverse fields through the off-diagonal term \offdiag \cite{chakraborty_theory_2004,dollberg_effect_2022}. While the macroscopic transverse demagnetization field throughout the entire sample will average out to zero in the bulk by symmetry, a finite variance means that any individual spin will experience a non-zero transverse demagnetization field locally, whose strength increases as the sample shape aspect ratio varies from an ideal needle to a regular cuboid. Thus, in our cuboid sample, the increased quantum fluctuations due to the shape-dependent transverse fields can drive up the crossover temperature from quantum to classical behavior, thereby keeping our sample deep in the quantum regime even at temperatures approaching the classical phase transition.

While the application of a modest external transverse field ($H_\perp = 200 \text{ Oe}$) has no observable impacts on the macroscopic sample response at slower timescales ($\sim 1 \text{ s}$), the situation changes dramatically when analyzing individual avalanches on a $\mu \text{s}$ timescale. The motion of individual domains on this timescale was measured through digitizing the voltage output of an inductive pickup coil measuring the time-derivative of the sample magnetization, $dM/dt$, recorded on an oscilloscope sampling at 1 MHz. Using a thresholding technique as shown in Fig. \ref{fig:2}c (see methods section for details), individual avalanches with voltages rising above the noise floor were extracted in software, and the statistics of their metrics (duration $T$ and area $S$ among others) were analyzed, along with cross-correlations between the various metrics according to traditional crackling noise analysis \cite{perkovic_avalanches_1995,zapperi_dynamics_1998}.

We note that the only events that we are able to observe correspond to the largest avalanches. The detectable avalanches range in size from $\approx 2 \times 10^{-10} \text{ Wb} \rightarrow 6 \times 10^{-9} \text{ Wb}$, corresponding to avalanches containing a number of spins $N$ between $1.5 \times 10^{15}$ and $4.5 \times 10^{16}$ spins. The change in the macroscopic magnetization over the full hysteresis loop (Fig. \ref{fig:2}a) is actually dominated by the multiplicity of smaller domain flips below our noise floor, with the change in magnetization due to the measured large events contributing anywhere from $\sim 0.01 \%$ (at $H_\perp = 0$) to $\sim 0.1 \%$ (at $H_\perp = 200$ Oe) of the total change in magnetization of the entire sample over the full loop. All events were observed in a narrow region between $150 \text{ Oe} \leq H_\parallel \leq 200 \text{ Oe})$ while ramping up, and at the equivalent negative field while ramping down (marked by arrows in Fig. \ref{fig:2}a). By contrast to previous susceptibility measurements \cite{brooke_tunable_2001} that measured only the domain walls with weakest pinning, in this experiment we measure only the most strongly-pinned domain walls.

While traditional Barkhausen analysis consists of deducing the underlying universality class by means of extracting the critical exponents through power-law fits of event statistics \cite{perkovic_avalanches_1995,zapperi_dynamics_1998} or by using lineshape analysis to learn about underlying dissipative or demagnetization effects \cite{zapperi_signature_2005,papanikolaou_universality_2011,durin_quantitative_2016,silevitch_magnetic_2019}, this is not appropriate here because our data does not display the standard power-laws characteristic of universality. Instead, we observe two distinct “classes” of events, presumably corresponding to two different domain-wall activation mechanisms, that show remarkably different dependences on applied transverse fields. 

\begin{figure}[tb]
    \centering
    \includegraphics[width=0.4\textwidth]{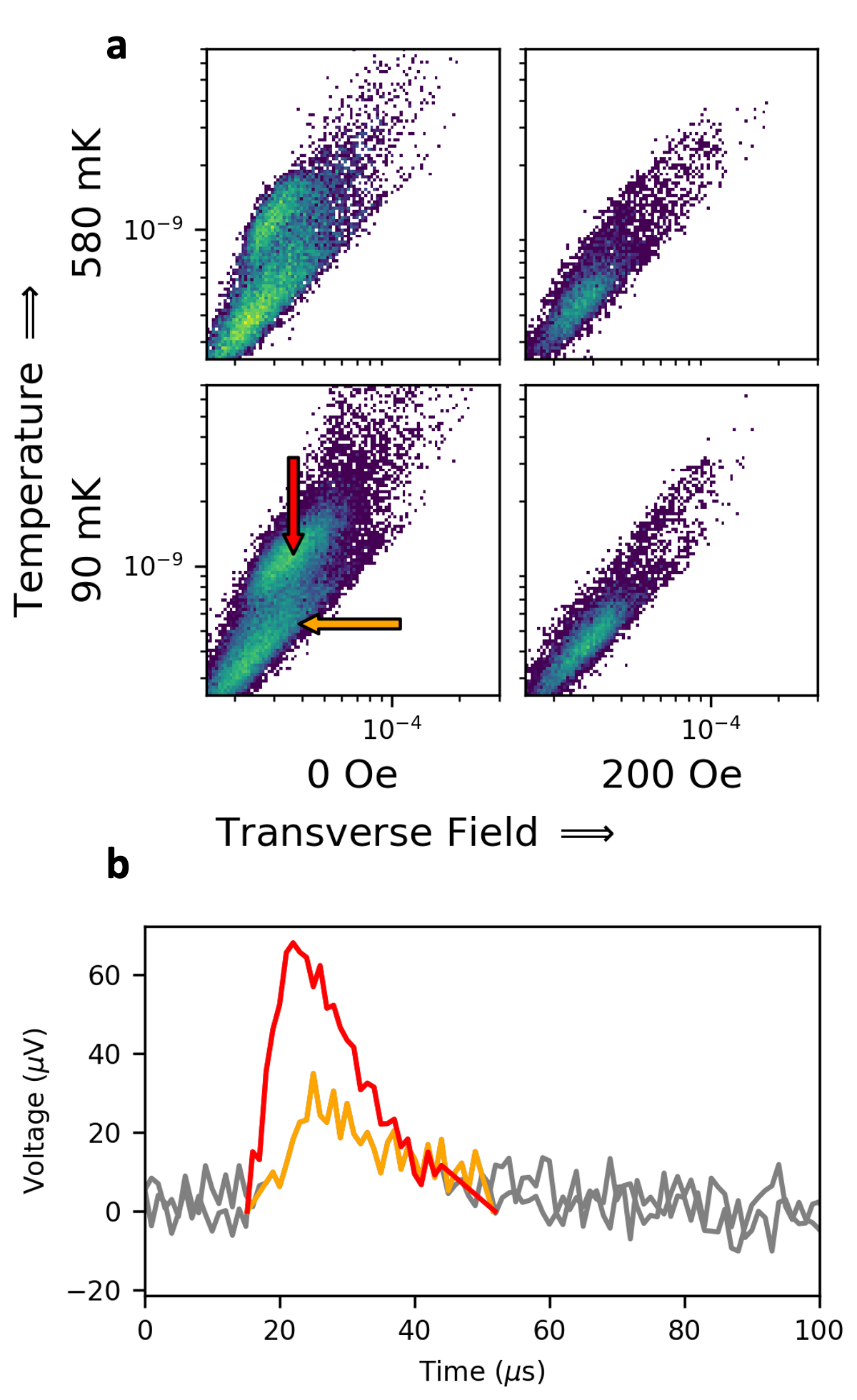}
    \caption{\textbf{Classes of events} \textbf{a} 2D histograms of event area (y-axis) vs event duration (x-axis) for low/high temperatures and transverse fields. \textbf{b} Sample events of each class: ``independent'' event in orange, and ``cooperative'' event in red as indicated by the colored arrows in (a)}
    \label{fig:3}
\end{figure}

The non-critical behavior is most easily observed by comparing the two-dimensional histograms plotting the cross-correlation between event duration ($T$) and area ($S = \int_0^T V dt$) for the most extreme temperatures (90 mK and 580 mK) and transverse fields (0 Oe and 200 Oe). As seen in Fig. \ref{fig:3}a, the events separate into two distinguishable classes at low fields: one class that we label as “independent” that approximately spans a power-law with an exponent of $\approx 1.1$ (close to the power of 1 indicative of avalanches \cite{papanikolaou_universality_2011}) over approximately one decade of duration, and the second that we designate “cooperative” (highlighted by the red oval in Fig. \ref{fig:3}a) that appears as an approximately Gaussian cluster over a more limited range of durations with higher areas for any given duration than events in the “independent” class. Furthermore, while the frequency of the “independent” events decreases only modestly with transverse field, the “cooperative” events are suppressed almost completely with a 200 Oe transverse field. We have plotted one sample event in each class in Fig. \ref{fig:3}b, both marked by arrows on the 2d histograms in Fig. \ref{fig:3}a, with the “cooperative” event in red and the “independent” event in orange.

\section{Discussion}

In the following discussion, we discuss the possible origins of these two activation mechanisms, as well as phenomenologically explain why such a small 200 Oe transverse field could suppress the “cooperative” events starting from the microscopic Hamiltonian.

First, we can deduce that both activation mechanisms are quantum mechanical in nature. Like the magnetization curves within the linear regime, the event statistics show the same temperature independence, demonstrating that within these experimental parameters, the sample is deep within the quantum regime, where the dynamics are governed by quantum tunneling, rather than thermal activation, of spins. 

Given that both activation mechanisms are due to quantum tunneling (rather than one being quantum and the other thermal), it is not immediately obvious how there could be two different tunneling mechanisms, why they would have such dramatically different transverse field dependence, or how such a small 200 Oe field could suppress markedly either class of events. 

Given this challenge, it is necessary to go beyond the theoretical picture of a single independent wall tunneling and consider the interaction between walls. In so doing, we can recover a phenomenological model in which the two different activation mechanisms correspond, on one hand, to walls tunneling independently of each other and, on the other, to cooperative tunneling of pairs of walls. Co-tunneling of domain walls is strongly affected by the application of an external transverse field much smaller than the fields required to induce single-spin tunneling.

We consider a model in which a single plane wall, or an adjacent pair of walls, can displace themselves through the system. A detailed consideration of this model is given in the supplementary information, but we summarize the relevant conclusions of the theoretical treatment below.

\begin{figure}[tb]
    \centering
    \includegraphics[width=0.4\textwidth]{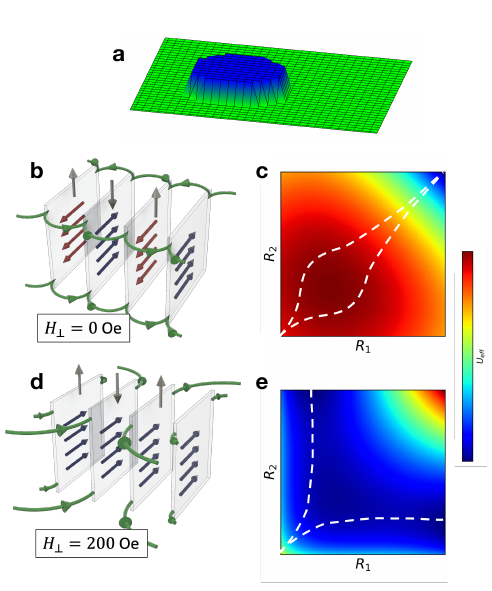}
    \caption{\textbf{Domain wall configurations and  interaction potentials.} \textbf{a} Schematic of Bloch domain wall with single ``plaquette'' structure. Grid denotes locations of individual spins. \textbf{b,d} Vertical grey arrows designate bulk magnetization direction within a domain along the Ising axis, while the red/blue arrows designate the transverse polarizations within a Bloch wall. The green curved arrows illustrate the demagnetization fields. The tunneling potentials \textbf{c,e} are a function of the radii of the two interacting plaquettes, $R_1$ and $R_2$, coupled via the dipolar interaction. \textbf{b} Staggered polarizations of the domain walls at zero transverse field with corresponding attractive interaction in \textbf{c} causing $R_1$ and $R_2$ to grow together (as indicated by the tunneling paths shown in grey). \textbf{d} All walls polarized in the same direction due to the transverse field, with the corresponding repulsive interaction in \textbf{e}, cause plaquettes to grow independently from one another, as indicated by the tunneling paths shown in white.}
    \label{fig:4}
\end{figure}

Even without disorder (for $x=1$), the walls are pinned by a lattice periodic potential. Wall displacement then occurs by the nucleation of plaquettes of a displaced wall (Fig.~\ref{fig:4}a). This occurs by tunneling through a barrier created by the line tension in the plaquette periphery and the tunneling is driven by an applied longitudinal field. Plaquettes can nucleate at different parts of a wall, as well as on top of each other; they are the 2-dimensional lattice version of quantum bubble nucleation \cite{nakamuraSinglecollectivedegreeoffreedomModelsMacroscopic1995}.

In addition to the independent tunneling processes of a single plaquette, there are co-tunneling processes involving the dipolar interaction between two plaquettes. The interaction term between two plaquettes, $\Delta U_{12}$, as a function of the radius of each plaquette, $R_i$, is given by:

\begin{equation}
    \Delta U_{12} = -\frac{\mu_0}{r_{12}^3} (g \mu_B \pi)^2 \left[  3 \frac{(\vec{\Tilde{J}}_1 \cdot \vec{r}_{12})(\vec{\Tilde{J}}_2 \cdot \vec{r}_{12})}{r_{12}^2} - \vec{\Tilde{J}}_1 \cdot \vec{\Tilde{J}}_2 \right] \frac{R_1^2R_2^2}{a_0^4}
\end{equation}

where $\vec{\Tilde{J}}_i$ is the spin per Ho ion in each plaquette and $\vec{r}_{12}$ is the separation between plaquettes. If this interaction term is attractive ($\Delta U_{12} < 0$), then the configuration energy will be minimized by having the radii of both plaquettes ($R_1,R_2$) growing together. In this picture, attractive interactions between plaquettes on different domain walls can cause co-tunneling processes in which nucleation of one plaquette lowers the energy barrier for nucleation of the other. By contrast, if the interaction term is repulsive ($\Delta U_{12} > 0$), then these co-tunneling processes are suppressed, and the energetically favored tunneling paths consist of each plaquette growing independently of the other.

The sign of this interaction depends on the relative orientation and polarization of the two plaquettes. If we assume that the plaquettes are opposite each other, and that their polarizations lie along the x-axis, the interaction simplifies to being attractive when the polarizations are aligned and being repulsive when they are anti-aligned. Hence, the polarizations of domain walls affect the tunneling dynamics by giving rise to cooperative tunneling of pairs of plaquettes on adjacent domain walls when their polarizations are anti-aligned, in addition to the standard independent tunneling processes of single plaquettes.

While the application of an external longitudinal field simply reduces the height of the tunneling barrier, accelerating plaquette nucleation and domain wall tunneling, application of a sufficiently strong transverse field will act to orient polarizations of all the walls to be parallel to the applied field. While the spatial extent of the measured avalanches is not known, the large number of spins involved in an avalanche ($\sim 10^{15}$--$10^{16}$) guarantees that that wall area will be large. Furthermore, since the Zeeman energy of the wall scales as the number of spins within the wall, the polarizations of these large sections of domain wall will be highly susceptible to even a modest transverse field. Thus, while a small 200 Oe transverse field is much less than the single-ion tunneling field scale ($\sim$ 20 kOe)\cite{schechter_significance_2005}, or the quantum phase transition (QPT) field scale ($\sim$ 12 kOe)\cite{silevitch_switchable_2010}, it is large enough to appreciably polarize most domain walls in the same direction, thereby changing the statistics of the co-tunneling processes. We illustrate these two wall configurations in Fig. \ref{fig:4}, with the wall polarizations staggered in zero transverse field in panel b, and the configuration with all walls polarized in the same direction in a finite transverse field in panel d.

When the external field is zero, the wall polarizations are random, giving rise to both aligned and anti-aligned domain wall configurations. Consequently, at zero field, co-tunneling processes will be possible due to the attractive interaction between the anti-aligned domain walls. When a finite transverse field is applied, however, all the walls polarize in the same direction, making all of the interactions between walls repulsive, and consequently suppressing the co-tunneling processes. We plot the corresponding tunneling potentials in Fig. \ref{fig:4}c,e for the domain wall configurations in panels (b,d). (c) illustrates the attractive interaction due to the staggered polarizations of Bloch walls in zero field;  (e) illustrates the repulsive interaction due to all wall polarizations being aligned along the external transverse field direction.

We emphasize that, while all domain wall motion is governed primarily through quantum tunneling of small plaquettes, large events consisting of $10^{15}$ spins are \textit{avalanches} of many plaquette tunneling events triggering one another--not single coherent macroscopic tunneling events. Co-tunneling processes do not cause entire walls to tunnel together, but they can cause imultaneous quantum nucleation of two plaquettes on adjacent walls at once, affecting the avalanche statistics through an introduction of a second avalanche activation mechanism.

\section{Conclusion}

We conclude by emphasizing the key results found here. We show that, within the linear regime of a hysteresis loop (dominated by domain wall motion, rather than domain creation/annihilation), the sample dynamics are completely temperature independent up to 95\% of $T_c$, indicating that all domain wall motion is governed by quantum tunneling, rather than thermal activation. Furthermore, we find that, in addition to independent activation of individual domain walls, our sample exhibits a second activation mechanism, where dipolar interactions between neighboring domain walls causes cooperative tunneling that simultaneously nucleates plaquettes on both domain walls, triggering correlated avalanches. These co-tunneling processes can be suppressed strongly with a modest transverse field--much weaker than any field scale governing either the single-spin tunneling rates or the many-body phase diagram.

These experiments involve domain wall motion in a ferromagnet, the prototype for Barkhausen noise measurements. Avalanches, however, occur in a diverse set of physical systems, from photomultiplier tubes, to earthquakes, to plastic deformation of nanostructures, revealing details of a system’s energy landscape and reversal dynamics. Our work, venturing into the quantum regime suggests that similar quantum effects should be observable in other systems where long-range interactions between microscopic degrees of freedom can cause correlated activation of macroscopic avalanches, i.e., that one should search for quantum avalanche and quantum Barkhausen effects in a large variety of systems.

\section{Methods}
A 5mm x 5mm x 10mm tetragonal crystal of LiHo\textsubscript{0.4}Y\textsubscript{0.6}F\textsubscript{4}, with long axis parallel to the Ising axis of the localized Ho\textsuperscript{3+} moments, was mounted in an insulating PEEK coil form to resist torques applied from the external transverse field. A 100-turn inductive pickup coil was wrapped around the center of the sample to measure the time-derivative of the bulk magnetization, and the coil assembly was mounted to the high-purity copper cold-finger of a Helium dilution refrigerator to reach milliKelvin temperature scales. An illustration of the experimental apparatus is included in Fig. \ref{fig:1}. Thermal equilibration was aided by heat-sinking the sample to an insulating sapphire plate, which was metallically-connected to the Cu cold-finger. External magnetic fields were applied using a superconducting 6T/2T vector magnet, with the longitudinal field ramped between saturation fields $\pm 4\: \textrm{kOe}$, with a sweep rate $dH_{\parallel}/dt = 11.1\: \textrm{Oe/s}$ in the adiabatic limit. In order to collect suitable statistics of the avalanche events, hundreds of hysteresis loops were taken at a series of temperatures from 90 mK (15\% $T_c$) $\rightarrow$ 580 mK (95\% $T_c$), and transverse fields from 0 Oe $\rightarrow$ 200 Oe. Despite this transverse field scale being an order of magnitude below a known transverse field scale corresponding to quantum speed-up in the same material, no avalanche events were observed above 200 Oe. We illustrate this by plotting the frequency of avalanche events per hysteresis loop as a function of transverse field in Fig. \ref{fig:2}b.

The quasistatic bulk magnetization of the sample was measured with the use of a GaAs Hall-effect magnetometer (Toshiba THS118) sampling at 1 Hz. We plot individual hysteresis loops at the extreme values of measured temperatures (90 mK and 580 mK) and transverse fields (0 Oe and 200 Oe) in Figure 1(c). There is little variation in the magnetization loops at the different temperatures, and there is no observable difference in sample response from 0 Oe to 200 Oe for any temperature. 

In order to measure individual avalanche events, a higher sampling frequency is needed than the 1 Hz sampling of the Hall sensor. Instead of observing avalanches through measurements of the magnetization, the inductive pickup coil convered the time-derivative of the magnetization into a voltage signal, which was further amplified by a 
high-frequency transformer amplifier (CMR-Direct LTT-h) with a gain of 10 to extend out the frequency response past the 100 kHz attempt frequency of LiHoF\textsubscript{4}. This voltage signal was further amplified at room temperature by a transistor pre-amplifier (Stanford Research SR560) with a gain of 5000 running off of DC battery power, and digitized at a sampling frequency of 1 MHz with an oscilloscope (PicoScope 4262). 

Individual avalanche events were extracted from the raw voltage traces using an automated routine that identified events as segments of the data with voltages greater than a threshold calculated to be 3.5$\sigma$ of the gaussian-distributed instrumentation noise, after subtracting off a low-frequency ($<$ 1 kHz) background. The beginning and ends of each individual event was linearly-extrapolated to 0 V that was independent of extrapolation regime. We plot a sample event extraction in Fig. \ref{fig:2}c.

\begin{acknowledgments}We thank R.D. McKenzie for illuminating discussions. The experimental work and the modeling of the data at Caltech were supported by the U.S. Department of Energy Basic Energy Sciences Award No. DE-SC0014866. Theoretical work at UBC was supported by the National Sciences and Engineering Research Council of Canada, No. RGPIN-2019-05582.
\end{acknowledgments}

Correspondence and requests for materials should be addressed to T.F.R., tfr@caltech.edu, and P.C.E.S, stamp@phas.ubc.ca. 

\bibliography{DW_Paper_Refs.bib}

\begin{thebibliography}{35}%
\makeatletter
\providecommand \@ifxundefined [1]{%
 \@ifx{#1\undefined}
}%
\providecommand \@ifnum [1]{%
 \ifnum #1\expandafter \@firstoftwo
 \else \expandafter \@secondoftwo
 \fi
}%
\providecommand \@ifx [1]{%
 \ifx #1\expandafter \@firstoftwo
 \else \expandafter \@secondoftwo
 \fi
}%
\providecommand \natexlab [1]{#1}%
\providecommand \enquote  [1]{``#1''}%
\providecommand \bibnamefont  [1]{#1}%
\providecommand \bibfnamefont [1]{#1}%
\providecommand \citenamefont [1]{#1}%
\providecommand \href@noop [0]{\@secondoftwo}%
\providecommand \href [0]{\begingroup \@sanitize@url \@href}%
\providecommand \@href[1]{\@@startlink{#1}\@@href}%
\providecommand \@@href[1]{\endgroup#1\@@endlink}%
\providecommand \@sanitize@url [0]{\catcode `\\12\catcode `\$12\catcode
  `\&12\catcode `\#12\catcode `\^12\catcode `\_12\catcode `\%12\relax}%
\providecommand \@@startlink[1]{}%
\providecommand \@@endlink[0]{}%
\providecommand \url  [0]{\begingroup\@sanitize@url \@url }%
\providecommand \@url [1]{\endgroup\@href {#1}{\urlprefix }}%
\providecommand \urlprefix  [0]{URL }%
\providecommand \Eprint [0]{\href }%
\providecommand \doibase [0]{https://doi.org/}%
\providecommand \selectlanguage [0]{\@gobble}%
\providecommand \bibinfo  [0]{\@secondoftwo}%
\providecommand \bibfield  [0]{\@secondoftwo}%
\providecommand \translation [1]{[#1]}%
\providecommand \BibitemOpen [0]{}%
\providecommand \bibitemStop [0]{}%
\providecommand \bibitemNoStop [0]{.\EOS\space}%
\providecommand \EOS [0]{\spacefactor3000\relax}%
\providecommand \BibitemShut  [1]{\csname bibitem#1\endcsname}%
\let\auto@bib@innerbib\@empty
\bibitem [{\citenamefont {Barkhausen}(1919)}]{barkhausen_notitle_1919}%
  \BibitemOpen
  \bibfield  {author} {\bibinfo {author} {\bibfnamefont {H.}~\bibnamefont
  {Barkhausen}},\ }\href@noop {} {\bibfield  {journal} {\bibinfo  {journal}
  {Physik Z.}\ }\textbf {\bibinfo {volume} {20}} (\bibinfo {year}
  {1919})}\BibitemShut {NoStop}%
\bibitem [{\citenamefont {Dahmen}\ and\ \citenamefont
  {Sethna}(1996)}]{dahmen_hysteresis_1996}%
  \BibitemOpen
  \bibfield  {author} {\bibinfo {author} {\bibfnamefont {K.}~\bibnamefont
  {Dahmen}}\ and\ \bibinfo {author} {\bibfnamefont {J.~P.}\ \bibnamefont
  {Sethna}},\ }\bibfield  {title} {\bibinfo {title} {Hysteresis, avalanches,
  and disorder-induced critical scaling: {A} renormalization-group approach},\
  }\href {https://doi.org/10.1103/PhysRevB.53.14872} {\bibfield  {journal}
  {\bibinfo  {journal} {Physical Review B}\ }\textbf {\bibinfo {volume} {53}},\
  \bibinfo {pages} {14872} (\bibinfo {year} {1996})},\ \bibinfo {note}
  {publisher: American Physical Society}\BibitemShut {NoStop}%
\bibitem [{\citenamefont {Alessandro}\ \emph {et~al.}(1990)\citenamefont
  {Alessandro}, \citenamefont {Beatrice}, \citenamefont {Bertotti},\ and\
  \citenamefont {Montorsi}}]{alessandro_domainwall_1990}%
  \BibitemOpen
  \bibfield  {author} {\bibinfo {author} {\bibfnamefont {B.}~\bibnamefont
  {Alessandro}}, \bibinfo {author} {\bibfnamefont {C.}~\bibnamefont
  {Beatrice}}, \bibinfo {author} {\bibfnamefont {G.}~\bibnamefont {Bertotti}},\
  and\ \bibinfo {author} {\bibfnamefont {A.}~\bibnamefont {Montorsi}},\
  }\bibfield  {title} {\bibinfo {title} {Domain‐wall dynamics and
  {Barkhausen} effect in metallic ferromagnetic materials. {I}. {Theory}},\
  }\href {https://doi.org/10.1063/1.346423} {\bibfield  {journal} {\bibinfo
  {journal} {Journal of Applied Physics}\ }\textbf {\bibinfo {volume} {68}},\
  \bibinfo {pages} {2901} (\bibinfo {year} {1990})}\BibitemShut {NoStop}%
\bibitem [{\citenamefont {Stamp}\ \emph {et~al.}(1992)\citenamefont {Stamp},
  \citenamefont {Chudnovsky},\ and\ \citenamefont
  {Barbara}}]{stamp_quantum_1992}%
  \BibitemOpen
  \bibfield  {author} {\bibinfo {author} {\bibfnamefont {P.}~\bibnamefont
  {Stamp}}, \bibinfo {author} {\bibfnamefont {E.}~\bibnamefont {Chudnovsky}},\
  and\ \bibinfo {author} {\bibfnamefont {B.}~\bibnamefont {Barbara}},\
  }\bibfield  {title} {\bibinfo {title} {{QUANTUM} {TUNNELING} {OF}
  {MAGNETIZATION} {IN} {SOLIDS}},\ }\href
  {https://doi.org/10.1142/S0217979292000670} {\bibfield  {journal} {\bibinfo
  {journal} {International Journal of Modern Physics B}\ }\textbf {\bibinfo
  {volume} {06}},\ \bibinfo {pages} {1355} (\bibinfo {year}
  {1992})}\BibitemShut {NoStop}%
\bibitem [{\citenamefont {Hong}\ and\ \citenamefont
  {Giordano}(1995)}]{hong_approach_1995}%
  \BibitemOpen
  \bibfield  {author} {\bibinfo {author} {\bibfnamefont {K.}~\bibnamefont
  {Hong}}\ and\ \bibinfo {author} {\bibfnamefont {N.}~\bibnamefont
  {Giordano}},\ }\bibfield  {title} {\bibinfo {title} {Approach to mesoscopic
  magnetic measurements},\ }\href {https://doi.org/10.1103/PhysRevB.51.9855}
  {\bibfield  {journal} {\bibinfo  {journal} {Physical Review B}\ }\textbf
  {\bibinfo {volume} {51}},\ \bibinfo {pages} {9855} (\bibinfo {year}
  {1995})},\ \bibinfo {note} {publisher: American Physical Society}\BibitemShut
  {NoStop}%
\bibitem [{\citenamefont {Hong}\ and\ \citenamefont
  {Giordano}(1996)}]{hong_effect_1996}%
  \BibitemOpen
  \bibfield  {author} {\bibinfo {author} {\bibfnamefont {K.}~\bibnamefont
  {Hong}}\ and\ \bibinfo {author} {\bibfnamefont {N.}~\bibnamefont
  {Giordano}},\ }\bibfield  {title} {\bibinfo {title} {Effect of microwaves on
  domain wall motion in thin {Ni} wires},\ }\href
  {https://doi.org/10.1209/epl/i1996-00201-y} {\bibfield  {journal} {\bibinfo
  {journal} {Europhysics Letters}\ }\textbf {\bibinfo {volume} {36}},\ \bibinfo
  {pages} {147} (\bibinfo {year} {1996})},\ \bibinfo {note} {publisher: IOP
  Publishing}\BibitemShut {NoStop}%
\bibitem [{\citenamefont {Wernsdorfer}\ \emph {et~al.}(2005)\citenamefont
  {Wernsdorfer}, \citenamefont {Clérac}, \citenamefont {Coulon}, \citenamefont
  {Lecren},\ and\ \citenamefont {Miyasaka}}]{wernsdorfer_quantum_2005}%
  \BibitemOpen
  \bibfield  {author} {\bibinfo {author} {\bibfnamefont {W.}~\bibnamefont
  {Wernsdorfer}}, \bibinfo {author} {\bibfnamefont {R.}~\bibnamefont
  {Clérac}}, \bibinfo {author} {\bibfnamefont {C.}~\bibnamefont {Coulon}},
  \bibinfo {author} {\bibfnamefont {L.}~\bibnamefont {Lecren}},\ and\ \bibinfo
  {author} {\bibfnamefont {H.}~\bibnamefont {Miyasaka}},\ }\bibfield  {title}
  {\bibinfo {title} {Quantum {Nucleation} in a {Single}-{Chain} {Magnet}},\
  }\href {https://doi.org/10.1103/PhysRevLett.95.237203} {\bibfield  {journal}
  {\bibinfo  {journal} {Physical Review Letters}\ }\textbf {\bibinfo {volume}
  {95}},\ \bibinfo {pages} {237203} (\bibinfo {year} {2005})},\ \bibinfo {note}
  {publisher: American Physical Society}\BibitemShut {NoStop}%
\bibitem [{\citenamefont {Brooke}\ \emph {et~al.}(2001)\citenamefont {Brooke},
  \citenamefont {Rosenbaum},\ and\ \citenamefont
  {Aeppli}}]{brooke_tunable_2001}%
  \BibitemOpen
  \bibfield  {author} {\bibinfo {author} {\bibfnamefont {J.}~\bibnamefont
  {Brooke}}, \bibinfo {author} {\bibfnamefont {T.~F.}\ \bibnamefont
  {Rosenbaum}},\ and\ \bibinfo {author} {\bibfnamefont {G.}~\bibnamefont
  {Aeppli}},\ }\bibfield  {title} {\bibinfo {title} {Tunable quantum tunnelling
  of magnetic domain walls},\ }\href {https://doi.org/10.1038/35098037}
  {\bibfield  {journal} {\bibinfo  {journal} {Nature}\ }\textbf {\bibinfo
  {volume} {413}},\ \bibinfo {pages} {610} (\bibinfo {year} {2001})},\ \bibinfo
  {note} {number: 6856 Publisher: Nature Publishing Group}\BibitemShut
  {NoStop}%
\bibitem [{\citenamefont {Silevitch}\ \emph {et~al.}(2010)\citenamefont
  {Silevitch}, \citenamefont {Aeppli},\ and\ \citenamefont
  {Rosenbaum}}]{silevitch_switchable_2010}%
  \BibitemOpen
  \bibfield  {author} {\bibinfo {author} {\bibfnamefont {D.~M.}\ \bibnamefont
  {Silevitch}}, \bibinfo {author} {\bibfnamefont {G.}~\bibnamefont {Aeppli}},\
  and\ \bibinfo {author} {\bibfnamefont {T.~F.}\ \bibnamefont {Rosenbaum}},\
  }\bibfield  {title} {\bibinfo {title} {Switchable hardening of a ferromagnet
  at fixed temperature},\ }\href {https://doi.org/10.1073/pnas.0910575107}
  {\bibfield  {journal} {\bibinfo  {journal} {Proceedings of the National
  Academy of Sciences}\ }\textbf {\bibinfo {volume} {107}},\ \bibinfo {pages}
  {2797} (\bibinfo {year} {2010})},\ \bibinfo {note} {publisher: Proceedings of
  the National Academy of Sciences}\BibitemShut {NoStop}%
\bibitem [{\citenamefont {Silevitch}\ \emph {et~al.}(2019)\citenamefont
  {Silevitch}, \citenamefont {Xu}, \citenamefont {Tang}, \citenamefont
  {Dahmen},\ and\ \citenamefont {Rosenbaum}}]{silevitch_magnetic_2019}%
  \BibitemOpen
  \bibfield  {author} {\bibinfo {author} {\bibfnamefont {D.~M.}\ \bibnamefont
  {Silevitch}}, \bibinfo {author} {\bibfnamefont {J.}~\bibnamefont {Xu}},
  \bibinfo {author} {\bibfnamefont {C.}~\bibnamefont {Tang}}, \bibinfo {author}
  {\bibfnamefont {K.~A.}\ \bibnamefont {Dahmen}},\ and\ \bibinfo {author}
  {\bibfnamefont {T.~F.}\ \bibnamefont {Rosenbaum}},\ }\bibfield  {title}
  {\bibinfo {title} {Magnetic domain dynamics in an insulating quantum
  ferromagnet},\ }\href {https://doi.org/10.1103/PhysRevB.100.134405}
  {\bibfield  {journal} {\bibinfo  {journal} {Physical Review B}\ }\textbf
  {\bibinfo {volume} {100}},\ \bibinfo {pages} {134405} (\bibinfo {year}
  {2019})},\ \bibinfo {note} {publisher: American Physical Society}\BibitemShut
  {NoStop}%
\bibitem [{\citenamefont {Tatara}\ and\ \citenamefont
  {Fukuyama}(1994{\natexlab{a}})}]{tatara_macroscopic_1994}%
  \BibitemOpen
  \bibfield  {author} {\bibinfo {author} {\bibfnamefont {G.}~\bibnamefont
  {Tatara}}\ and\ \bibinfo {author} {\bibfnamefont {H.}~\bibnamefont
  {Fukuyama}},\ }\bibfield  {title} {\bibinfo {title} {Macroscopic {Quantum}
  {Tunneling} of a {Domain} {Wall} in a {Ferromagnetic} {Metal}},\ }\href
  {https://doi.org/10.1143/jpsj.63.2538} {\bibfield  {journal} {\bibinfo
  {journal} {Journal of the Physical Society of Japan}\ }\textbf {\bibinfo
  {volume} {63}},\ \bibinfo {pages} {2538} (\bibinfo {year}
  {1994}{\natexlab{a}})}\BibitemShut {NoStop}%
\bibitem [{\citenamefont {Tatara}\ and\ \citenamefont
  {Fukuyama}(1994{\natexlab{b}})}]{tatara_macroscopic_1994-1}%
  \BibitemOpen
  \bibfield  {author} {\bibinfo {author} {\bibfnamefont {G.}~\bibnamefont
  {Tatara}}\ and\ \bibinfo {author} {\bibfnamefont {H.}~\bibnamefont
  {Fukuyama}},\ }\bibfield  {title} {\bibinfo {title} {Macroscopic quantum
  tunneling of a domain wall in a ferromagnetic metal},\ }\href
  {https://doi.org/10.1103/PhysRevLett.72.772} {\bibfield  {journal} {\bibinfo
  {journal} {Physical Review Letters}\ }\textbf {\bibinfo {volume} {72}},\
  \bibinfo {pages} {772} (\bibinfo {year} {1994}{\natexlab{b}})},\ \bibinfo
  {note} {publisher: American Physical Society}\BibitemShut {NoStop}%
\bibitem [{\citenamefont {Stamp}(1991)}]{stamp_quantum_1991}%
  \BibitemOpen
  \bibfield  {author} {\bibinfo {author} {\bibfnamefont {P.~C.~E.}\
  \bibnamefont {Stamp}},\ }\bibfield  {title} {\bibinfo {title} {Quantum
  dynamics and tunneling of domain walls in ferromagnetic insulators},\ }\href
  {https://doi.org/10.1103/PhysRevLett.66.2802} {\bibfield  {journal} {\bibinfo
   {journal} {Physical Review Letters}\ }\textbf {\bibinfo {volume} {66}},\
  \bibinfo {pages} {2802} (\bibinfo {year} {1991})},\ \bibinfo {note}
  {publisher: American Physical Society}\BibitemShut {NoStop}%
\bibitem [{\citenamefont {Dubé}\ and\ \citenamefont
  {Stamp}(1998)}]{dube_effects_1998}%
  \BibitemOpen
  \bibfield  {author} {\bibinfo {author} {\bibfnamefont {M.}~\bibnamefont
  {Dubé}}\ and\ \bibinfo {author} {\bibfnamefont {P.~C.~E.}\ \bibnamefont
  {Stamp}},\ }\bibfield  {title} {\bibinfo {title} {Effects of {Phonons} and
  {Nuclear} {Spins} on the {Tunneling} of a {Domain} {Wall}},\ }\href
  {https://doi.org/10.1023/A:1022676810365} {\bibfield  {journal} {\bibinfo
  {journal} {Journal of Low Temperature Physics}\ }\textbf {\bibinfo {volume}
  {110}},\ \bibinfo {pages} {779} (\bibinfo {year} {1998})}\BibitemShut
  {NoStop}%
\bibitem [{\citenamefont {Takagi}\ and\ \citenamefont
  {Tatara}(1996)}]{takagi_macroscopic_1996}%
  \BibitemOpen
  \bibfield  {author} {\bibinfo {author} {\bibfnamefont {S.}~\bibnamefont
  {Takagi}}\ and\ \bibinfo {author} {\bibfnamefont {G.}~\bibnamefont
  {Tatara}},\ }\bibfield  {title} {\bibinfo {title} {Macroscopic quantum
  coherence of chirality of a domain wall in ferromagnets},\ }\href
  {https://doi.org/10.1103/PhysRevB.54.9920} {\bibfield  {journal} {\bibinfo
  {journal} {Physical Review B}\ }\textbf {\bibinfo {volume} {54}},\ \bibinfo
  {pages} {9920} (\bibinfo {year} {1996})},\ \bibinfo {note} {publisher:
  American Physical Society}\BibitemShut {NoStop}%
\bibitem [{\citenamefont {Galkina}\ \emph {et~al.}(2008)\citenamefont
  {Galkina}, \citenamefont {Ivanov}, \citenamefont {Savel’ev},\ and\
  \citenamefont {Nori}}]{galkina_chirality_2008}%
  \BibitemOpen
  \bibfield  {author} {\bibinfo {author} {\bibfnamefont {E.~G.}\ \bibnamefont
  {Galkina}}, \bibinfo {author} {\bibfnamefont {B.~A.}\ \bibnamefont {Ivanov}},
  \bibinfo {author} {\bibfnamefont {S.}~\bibnamefont {Savel’ev}},\ and\
  \bibinfo {author} {\bibfnamefont {F.}~\bibnamefont {Nori}},\ }\bibfield
  {title} {\bibinfo {title} {Chirality tunneling and quantum dynamics for
  domain walls in mesoscopic ferromagnets},\ }\href
  {https://doi.org/10.1103/PhysRevB.77.134425} {\bibfield  {journal} {\bibinfo
  {journal} {Physical Review B}\ }\textbf {\bibinfo {volume} {77}},\ \bibinfo
  {pages} {134425} (\bibinfo {year} {2008})},\ \bibinfo {note} {publisher:
  American Physical Society}\BibitemShut {NoStop}%
\bibitem [{\citenamefont {Durin}\ and\ \citenamefont
  {Zapperi}(2006)}]{durin_barkhausen_2006}%
  \BibitemOpen
  \bibfield  {author} {\bibinfo {author} {\bibfnamefont {G.}~\bibnamefont
  {Durin}}\ and\ \bibinfo {author} {\bibfnamefont {S.}~\bibnamefont
  {Zapperi}},\ }\href@noop {} {\emph {\bibinfo {title} {The {Science} of
  {Hysteresis}}}},\ Vol.~\bibinfo {volume} {2}\ (\bibinfo  {publisher}
  {Elsevier},\ \bibinfo {year} {2006})\ pp.\ \bibinfo {pages}
  {181--267}\BibitemShut {NoStop}%
\bibitem [{\citenamefont {Bohn}\ \emph {et~al.}(2018)\citenamefont {Bohn},
  \citenamefont {Durin}, \citenamefont {Correa}, \citenamefont {Machado},
  \citenamefont {Della~Pace}, \citenamefont {Chesman},\ and\ \citenamefont
  {Sommer}}]{bohn_playing_2018}%
  \BibitemOpen
  \bibfield  {author} {\bibinfo {author} {\bibfnamefont {F.}~\bibnamefont
  {Bohn}}, \bibinfo {author} {\bibfnamefont {G.}~\bibnamefont {Durin}},
  \bibinfo {author} {\bibfnamefont {M.~A.}\ \bibnamefont {Correa}}, \bibinfo
  {author} {\bibfnamefont {N.~R.}\ \bibnamefont {Machado}}, \bibinfo {author}
  {\bibfnamefont {R.~D.}\ \bibnamefont {Della~Pace}}, \bibinfo {author}
  {\bibfnamefont {C.}~\bibnamefont {Chesman}},\ and\ \bibinfo {author}
  {\bibfnamefont {R.~L.}\ \bibnamefont {Sommer}},\ }\bibfield  {title}
  {\bibinfo {title} {Playing with universality classes of {Barkhausen}
  avalanches},\ }\href {https://doi.org/10.1038/s41598-018-29576-3} {\bibfield
  {journal} {\bibinfo  {journal} {Scientific Reports}\ }\textbf {\bibinfo
  {volume} {8}},\ \bibinfo {pages} {11294} (\bibinfo {year} {2018})},\ \bibinfo
  {note} {number: 1 Publisher: Nature Publishing Group}\BibitemShut {NoStop}%
\bibitem [{\citenamefont {de~Sousa}\ \emph {et~al.}(2020)\citenamefont
  {de~Sousa}, \citenamefont {dos Santos~Lima}, \citenamefont {Correa},
  \citenamefont {Sommer}, \citenamefont {Corso},\ and\ \citenamefont
  {Bohn}}]{de_sousa_waiting-time_2020}%
  \BibitemOpen
  \bibfield  {author} {\bibinfo {author} {\bibfnamefont {I.~P.}\ \bibnamefont
  {de~Sousa}}, \bibinfo {author} {\bibfnamefont {G.~Z.}\ \bibnamefont {dos
  Santos~Lima}}, \bibinfo {author} {\bibfnamefont {M.~A.}\ \bibnamefont
  {Correa}}, \bibinfo {author} {\bibfnamefont {R.~L.}\ \bibnamefont {Sommer}},
  \bibinfo {author} {\bibfnamefont {G.}~\bibnamefont {Corso}},\ and\ \bibinfo
  {author} {\bibfnamefont {F.}~\bibnamefont {Bohn}},\ }\bibfield  {title}
  {\bibinfo {title} {Waiting-time statistics in magnetic systems},\ }\href
  {https://doi.org/10.1038/s41598-020-66727-x} {\bibfield  {journal} {\bibinfo
  {journal} {Scientific Reports}\ }\textbf {\bibinfo {volume} {10}},\ \bibinfo
  {pages} {9692} (\bibinfo {year} {2020})},\ \bibinfo {note} {number: 1
  Publisher: Nature Publishing Group}\BibitemShut {NoStop}%
\bibitem [{\citenamefont {Schechter}\ and\ \citenamefont
  {Stamp}(2005)}]{schechter_significance_2005}%
  \BibitemOpen
  \bibfield  {author} {\bibinfo {author} {\bibfnamefont {M.}~\bibnamefont
  {Schechter}}\ and\ \bibinfo {author} {\bibfnamefont {P.~C.~E.}\ \bibnamefont
  {Stamp}},\ }\bibfield  {title} {\bibinfo {title} {Significance of the
  {Hyperfine} {Interactions} in the {Phase} {Diagram} of
  {LiHo$_{x}$Y$_{1-x}$F$_4$}},\ }\href
  {https://doi.org/10.1103/PhysRevLett.95.267208} {\bibfield  {journal}
  {\bibinfo  {journal} {Physical Review Letters}\ }\textbf {\bibinfo {volume}
  {95}},\ \bibinfo {pages} {267208} (\bibinfo {year} {2005})},\ \bibinfo {note}
  {publisher: American Physical Society}\BibitemShut {NoStop}%
\bibitem [{\citenamefont {Schechter}\ and\ \citenamefont
  {Stamp}(2008)}]{schechter_derivation_2008}%
  \BibitemOpen
  \bibfield  {author} {\bibinfo {author} {\bibfnamefont {M.}~\bibnamefont
  {Schechter}}\ and\ \bibinfo {author} {\bibfnamefont {P.~C.~E.}\ \bibnamefont
  {Stamp}},\ }\bibfield  {title} {\bibinfo {title} {Derivation of the
  low-\${T}\$ phase diagram of {LiHo$_{x}$Y$_{1-x}$F$_4$}: {A} dipolar quantum
  {Ising} magnet},\ }\href {https://doi.org/10.1103/PhysRevB.78.054438}
  {\bibfield  {journal} {\bibinfo  {journal} {Physical Review B}\ }\textbf
  {\bibinfo {volume} {78}},\ \bibinfo {pages} {054438} (\bibinfo {year}
  {2008})},\ \bibinfo {note} {publisher: American Physical Society}\BibitemShut
  {NoStop}%
\bibitem [{\citenamefont {Schechter}\ and\ \citenamefont
  {Laflorencie}(2006)}]{schechter_quantum_2006}%
  \BibitemOpen
  \bibfield  {author} {\bibinfo {author} {\bibfnamefont {M.}~\bibnamefont
  {Schechter}}\ and\ \bibinfo {author} {\bibfnamefont {N.}~\bibnamefont
  {Laflorencie}},\ }\bibfield  {title} {\bibinfo {title} {Quantum {Spin}
  {Glass} and the {Dipolar} {Interaction}},\ }\href
  {https://doi.org/10.1103/PhysRevLett.97.137204} {\bibfield  {journal}
  {\bibinfo  {journal} {Physical Review Letters}\ }\textbf {\bibinfo {volume}
  {97}},\ \bibinfo {pages} {137204} (\bibinfo {year} {2006})},\ \bibinfo {note}
  {publisher: American Physical Society}\BibitemShut {NoStop}%
\bibitem [{\citenamefont {Egami}(1973{\natexlab{a}})}]{egami_theory_1973-1}%
  \BibitemOpen
  \bibfield  {author} {\bibinfo {author} {\bibfnamefont {T.}~\bibnamefont
  {Egami}},\ }\bibfield  {title} {\bibinfo {title} {Theory of intrinsic
  magnetic after-effect i. thermally activated process},\ }\href
  {https://doi.org/10.1002/pssa.2210190242} {\bibfield  {journal} {\bibinfo
  {journal} {physica status solidi (a)}\ }\textbf {\bibinfo {volume} {19}},\
  \bibinfo {pages} {747} (\bibinfo {year} {1973}{\natexlab{a}})},\ \bibinfo
  {note} {\_eprint:
  https://onlinelibrary.wiley.com/doi/pdf/10.1002/pssa.2210190242}\BibitemShut
  {NoStop}%
\bibitem [{\citenamefont {Egami}(1973{\natexlab{b}})}]{egami_theory_1973}%
  \BibitemOpen
  \bibfield  {author} {\bibinfo {author} {\bibfnamefont {T.}~\bibnamefont
  {Egami}},\ }\bibfield  {title} {\bibinfo {title} {Theory of intrinsic
  magneitc after-effect {II}. {Tunnelling} process and comparison with
  experiments},\ }\href {https://doi.org/10.1002/pssa.2210200114} {\bibfield
  {journal} {\bibinfo  {journal} {physica status solidi (a)}\ }\textbf
  {\bibinfo {volume} {20}},\ \bibinfo {pages} {157} (\bibinfo {year}
  {1973}{\natexlab{b}})},\ \bibinfo {note} {\_eprint:
  https://onlinelibrary.wiley.com/doi/pdf/10.1002/pssa.2210200114}\BibitemShut
  {NoStop}%
\bibitem [{\citenamefont {Jorba}(2014)}]{jorba_shpm_2014}%
  \BibitemOpen
  \bibfield  {author} {\bibinfo {author} {\bibfnamefont {P.}~\bibnamefont
  {Jorba}},\ }\emph {\bibinfo {title} {{SHPM} imaging of {LiHoF$_4$} at ultra
  low temperatures}},\ \href@noop {} {Master's thesis},\ \bibinfo  {school}
  {Ecole Polytechnique Federale de Lausanne}, \bibinfo {address} {{Lausanne,
  Switzerland}} (\bibinfo {year} {2014})\BibitemShut {NoStop}%
\bibitem [{\citenamefont {Meyer}\ \emph {et~al.}(1989)\citenamefont {Meyer},
  \citenamefont {Pommier},\ and\ \citenamefont
  {Ferre}}]{meyer_magnetooptic_1989}%
  \BibitemOpen
  \bibfield  {author} {\bibinfo {author} {\bibfnamefont {P.}~\bibnamefont
  {Meyer}}, \bibinfo {author} {\bibfnamefont {J.}~\bibnamefont {Pommier}},\
  and\ \bibinfo {author} {\bibfnamefont {J.}~\bibnamefont {Ferre}},\ }\bibfield
   {title} {\bibinfo {title} {Magnetooptic {Observation} {Of} {Domains} {At}
  {Low} {Temperature} {In} {The} {Transparent} {Ferromagnet} {LiHoF}$_4$},\
  }\bibfield  {booktitle} {\emph {\bibinfo {booktitle} {Electro-{Optic} and
  {Magneto}-{Optic} {Materials} and {Applications}}},\ }\href
  {https://doi.org/10.1117/12.961386} {\bibfield  {journal} {\bibinfo
  {journal} {Electro-{{Optic}} and {{Magneto-Optic Materials}} and
  {{Applications}}}\ }\textbf {\bibinfo {volume} {1126}},\ \bibinfo {pages}
  {93} (\bibinfo {year} {1989})}\BibitemShut {NoStop}%
\bibitem [{\citenamefont {Herpin}(1968)}]{herpin_theorie_nodate}%
  \BibitemOpen
  \bibfield  {author} {\bibinfo {author} {\bibfnamefont {A.}~\bibnamefont
  {Herpin}},\ }\href {https://cir.nii.ac.jp/crid/1130282268992272128} {\emph
  {\bibinfo {title} {Théorie du magnétisme}}},\ Bibliothèque des sciences et
  techniques nucléaires\ (\bibinfo  {publisher} {Institut National des
  Sciences et Techniques Nucléaires Presses universitaires de France},\
  \bibinfo {year} {1968})\BibitemShut {NoStop}%
\bibitem [{\citenamefont {Chakraborty}\ \emph {et~al.}(2004)\citenamefont
  {Chakraborty}, \citenamefont {Henelius}, \citenamefont {Kjønsberg},
  \citenamefont {Sandvik},\ and\ \citenamefont
  {Girvin}}]{chakraborty_theory_2004}%
  \BibitemOpen
  \bibfield  {author} {\bibinfo {author} {\bibfnamefont {P.~B.}\ \bibnamefont
  {Chakraborty}}, \bibinfo {author} {\bibfnamefont {P.}~\bibnamefont
  {Henelius}}, \bibinfo {author} {\bibfnamefont {H.}~\bibnamefont
  {Kjønsberg}}, \bibinfo {author} {\bibfnamefont {A.~W.}\ \bibnamefont
  {Sandvik}},\ and\ \bibinfo {author} {\bibfnamefont {S.~M.}\ \bibnamefont
  {Girvin}},\ }\bibfield  {title} {\bibinfo {title} {Theory of the magnetic
  phase diagram of {LiHoF$_4$}},\ }\href
  {https://doi.org/10.1103/PhysRevB.70.144411} {\bibfield  {journal} {\bibinfo
  {journal} {Physical Review B}\ }\textbf {\bibinfo {volume} {70}},\ \bibinfo
  {pages} {144411} (\bibinfo {year} {2004})}\BibitemShut {NoStop}%
\bibitem [{\citenamefont {Dollberg}\ \emph {et~al.}(2022)\citenamefont
  {Dollberg}, \citenamefont {Andresen},\ and\ \citenamefont
  {Schechter}}]{dollberg_effect_2022}%
  \BibitemOpen
  \bibfield  {author} {\bibinfo {author} {\bibfnamefont {T.}~\bibnamefont
  {Dollberg}}, \bibinfo {author} {\bibfnamefont {J.~C.}\ \bibnamefont
  {Andresen}},\ and\ \bibinfo {author} {\bibfnamefont {M.}~\bibnamefont
  {Schechter}},\ }\bibfield  {title} {\bibinfo {title} {Effect of intrinsic
  quantum fluctuations on the phase diagram of anisotropic dipolar magnets},\
  }\href {https://doi.org/10.1103/PhysRevB.105.L180413} {\bibfield  {journal}
  {\bibinfo  {journal} {Physical Review B}\ }\textbf {\bibinfo {volume}
  {105}},\ \bibinfo {pages} {L180413} (\bibinfo {year} {2022})},\ \bibinfo
  {note} {publisher: American Physical Society}\BibitemShut {NoStop}%
\bibitem [{\citenamefont {Perković}\ \emph {et~al.}(1995)\citenamefont
  {Perković}, \citenamefont {Dahmen},\ and\ \citenamefont
  {Sethna}}]{perkovic_avalanches_1995}%
  \BibitemOpen
  \bibfield  {author} {\bibinfo {author} {\bibfnamefont {O.}~\bibnamefont
  {Perković}}, \bibinfo {author} {\bibfnamefont {K.}~\bibnamefont {Dahmen}},\
  and\ \bibinfo {author} {\bibfnamefont {J.~P.}\ \bibnamefont {Sethna}},\
  }\bibfield  {title} {\bibinfo {title} {Avalanches, {Barkhausen} {Noise}, and
  {Plain} {Old} {Criticality}},\ }\href
  {https://doi.org/10.1103/PhysRevLett.75.4528} {\bibfield  {journal} {\bibinfo
   {journal} {Physical Review Letters}\ }\textbf {\bibinfo {volume} {75}},\
  \bibinfo {pages} {4528} (\bibinfo {year} {1995})},\ \bibinfo {note}
  {publisher: American Physical Society}\BibitemShut {NoStop}%
\bibitem [{\citenamefont {Zapperi}\ \emph {et~al.}(1998)\citenamefont
  {Zapperi}, \citenamefont {Cizeau}, \citenamefont {Durin},\ and\ \citenamefont
  {Stanley}}]{zapperi_dynamics_1998}%
  \BibitemOpen
  \bibfield  {author} {\bibinfo {author} {\bibfnamefont {S.}~\bibnamefont
  {Zapperi}}, \bibinfo {author} {\bibfnamefont {P.}~\bibnamefont {Cizeau}},
  \bibinfo {author} {\bibfnamefont {G.}~\bibnamefont {Durin}},\ and\ \bibinfo
  {author} {\bibfnamefont {H.~E.}\ \bibnamefont {Stanley}},\ }\bibfield
  {title} {\bibinfo {title} {Dynamics of a ferromagnetic domain wall:
  {Avalanches}, depinning transition, and the {Barkhausen} effect},\ }\href
  {https://doi.org/10.1103/PhysRevB.58.6353} {\bibfield  {journal} {\bibinfo
  {journal} {Physical Review B}\ }\textbf {\bibinfo {volume} {58}},\ \bibinfo
  {pages} {6353} (\bibinfo {year} {1998})},\ \bibinfo {note} {publisher:
  American Physical Society}\BibitemShut {NoStop}%
\bibitem [{\citenamefont {Zapperi}\ \emph {et~al.}(2005)\citenamefont
  {Zapperi}, \citenamefont {Castellano}, \citenamefont {Colaiori},\ and\
  \citenamefont {Durin}}]{zapperi_signature_2005}%
  \BibitemOpen
  \bibfield  {author} {\bibinfo {author} {\bibfnamefont {S.}~\bibnamefont
  {Zapperi}}, \bibinfo {author} {\bibfnamefont {C.}~\bibnamefont {Castellano}},
  \bibinfo {author} {\bibfnamefont {F.}~\bibnamefont {Colaiori}},\ and\
  \bibinfo {author} {\bibfnamefont {G.}~\bibnamefont {Durin}},\ }\bibfield
  {title} {\bibinfo {title} {Signature of effective mass in crackling-noise
  asymmetry},\ }\href {https://doi.org/10.1038/nphys101} {\bibfield  {journal}
  {\bibinfo  {journal} {Nature Physics}\ }\textbf {\bibinfo {volume} {1}},\
  \bibinfo {pages} {46} (\bibinfo {year} {2005})},\ \bibinfo {note} {number: 1
  Publisher: Nature Publishing Group}\BibitemShut {NoStop}%
\bibitem [{\citenamefont {Papanikolaou}\ \emph {et~al.}(2011)\citenamefont
  {Papanikolaou}, \citenamefont {Bohn}, \citenamefont {Sommer}, \citenamefont
  {Durin}, \citenamefont {Zapperi},\ and\ \citenamefont
  {Sethna}}]{papanikolaou_universality_2011}%
  \BibitemOpen
  \bibfield  {author} {\bibinfo {author} {\bibfnamefont {S.}~\bibnamefont
  {Papanikolaou}}, \bibinfo {author} {\bibfnamefont {F.}~\bibnamefont {Bohn}},
  \bibinfo {author} {\bibfnamefont {R.~L.}\ \bibnamefont {Sommer}}, \bibinfo
  {author} {\bibfnamefont {G.}~\bibnamefont {Durin}}, \bibinfo {author}
  {\bibfnamefont {S.}~\bibnamefont {Zapperi}},\ and\ \bibinfo {author}
  {\bibfnamefont {J.~P.}\ \bibnamefont {Sethna}},\ }\bibfield  {title}
  {\bibinfo {title} {Universality beyond power laws and the average avalanche
  shape},\ }\href {https://doi.org/10.1038/nphys1884} {\bibfield  {journal}
  {\bibinfo  {journal} {Nature Physics}\ }\textbf {\bibinfo {volume} {7}},\
  \bibinfo {pages} {316} (\bibinfo {year} {2011})},\ \bibinfo {note} {number: 4
  Publisher: Nature Publishing Group}\BibitemShut {NoStop}%
\bibitem [{\citenamefont {Durin}\ \emph {et~al.}(2016)\citenamefont {Durin},
  \citenamefont {Bohn}, \citenamefont {Corrêa}, \citenamefont {Sommer},
  \citenamefont {Le~Doussal},\ and\ \citenamefont
  {Wiese}}]{durin_quantitative_2016}%
  \BibitemOpen
  \bibfield  {author} {\bibinfo {author} {\bibfnamefont {G.}~\bibnamefont
  {Durin}}, \bibinfo {author} {\bibfnamefont {F.}~\bibnamefont {Bohn}},
  \bibinfo {author} {\bibfnamefont {M.}~\bibnamefont {Corrêa}}, \bibinfo
  {author} {\bibfnamefont {R.}~\bibnamefont {Sommer}}, \bibinfo {author}
  {\bibfnamefont {P.}~\bibnamefont {Le~Doussal}},\ and\ \bibinfo {author}
  {\bibfnamefont {K.}~\bibnamefont {Wiese}},\ }\bibfield  {title} {\bibinfo
  {title} {Quantitative {Scaling} of {Magnetic} {Avalanches}},\ }\href
  {https://doi.org/10.1103/PhysRevLett.117.087201} {\bibfield  {journal}
  {\bibinfo  {journal} {Physical Review Letters}\ }\textbf {\bibinfo {volume}
  {117}},\ \bibinfo {pages} {087201} (\bibinfo {year} {2016})},\ \bibinfo
  {note} {publisher: American Physical Society}\BibitemShut {NoStop}%
\bibitem [{\citenamefont {Nakamura}\ \emph {et~al.}(1995)\citenamefont
  {Nakamura}, \citenamefont {Kanno},\ and\ \citenamefont
  {Takagi}}]{nakamuraSinglecollectivedegreeoffreedomModelsMacroscopic1995}%
  \BibitemOpen
  \bibfield  {author} {\bibinfo {author} {\bibfnamefont {T.}~\bibnamefont
  {Nakamura}}, \bibinfo {author} {\bibfnamefont {Y.}~\bibnamefont {Kanno}},\
  and\ \bibinfo {author} {\bibfnamefont {S.}~\bibnamefont {Takagi}},\
  }\bibfield  {title} {\bibinfo {title} {Single-collective-degree-of-freedom
  models of macroscopic quantum nucleation},\ }\href
  {https://doi.org/10.1103/PhysRevB.51.8446} {\bibfield  {journal} {\bibinfo
  {journal} {Physical Review B}\ }\textbf {\bibinfo {volume} {51}},\ \bibinfo
  {pages} {8446} (\bibinfo {year} {1995})}\BibitemShut {NoStop}%
\end{thebibliography}%

\end{document}